\documentstyle[psfig]{mn}

\newif\ifAMStwofonts

\def\msun{{\rm M_{\odot}}}

\def\be{\begin{equation}}
\def\ee{\end{equation}}
\def\x{XTE\thinspace J1118+480}
\def\xs{XTE\thinspace J1118+480 }

\ifoldfss
  \ifCUPmtlplainloaded \else
    \NewTextAlphabet{textbfit} {cmbxti10} {}
    \NewTextAlphabet{textbfss} {cmssbx10} {}
    \NewMathAlphabet{mathbfit} {cmbxti10} {} 
    \NewMathAlphabet{mathbfss} {cmssbx10} {} 
  \fi
  \ifAMStwofonts
    \ifCUPmtlplainloaded \else
      \NewSymbolFont{upmath} {eurm10}
      \NewSymbolFont{AMSa} {msam10}
      \NewMathSymbol{\upi}     {0}{upmath}{19}
      \NewMathSymbol{\umu}     {0}{upmath}{16}
      \NewMathSymbol{\upartial}{0}{upmath}{40}
      \NewMathSymbol{\leqslant}{3}{AMSa}{36}
      \NewMathSymbol{\geqslant}{3}{AMSa}{3E}

    \fi
  \fi
\fi 

\ifnfssone
  \newmathalphabet{\mathit}
  \addtoversion{normal}{\mathit}{cmr}{m}{it}
  \addtoversion{bold}{\mathit}{cmr}{bx}{it}
  \newmathalphabet{\mathbfit} 
  \addtoversion{normal}{\mathbfit}{cmr}{bx}{it}
  \addtoversion{bold}{\mathbfit}{cmr}{bx}{it}
  \newmathalphabet{\mathbfss} 
  \addtoversion{normal}{\mathbfss}{cmss}{bx}{n}
  \addtoversion{bold}{\mathbfss}{cmss}{bx}{n}
  \ifAMStwofonts
    \ifCUPmtlplainloaded \else
      \UseAMStwoboldmath
      \makeatletter
      \new@mathgroup\upmath@group
      \define@mathgroup\mv@normal\upmath@group{eur}{m}{n}
      \define@mathgroup\mv@bold\upmath@group{eur}{b}{n}
      \edef\UPM{\hexnumber\upmath@group}
      \new@mathgroup\amsa@group
      \define@mathgroup\mv@normal\amsa@group{msa}{m}{n}
      \define@mathgroup\mv@bold\amsa@group{msa}{m}{n}
      \edef\AMSa{\hexnumber\amsa@group}
      \makeatother
      \mathchardef\upi="0\UPM19
      \mathchardef\umu="0\UPM16
      \mathchardef\upartial="0\UPM40
      \mathchardef\leqslant="3\AMSa36
      \mathchardef\geqslant="3\AMSa3E
    \fi
  \fi
\fi 

\ifnfsstwo
  \DeclareMathAlphabet{\mathbfit}{OT1}{cmr}{bx}{it}
  \SetMathAlphabet\mathbfit{bold}{OT1}{cmr}{bx}{it}
  \DeclareMathAlphabet{\mathbfss}{OT1}{cmss}{bx}{n}
  \SetMathAlphabet\mathbfss{bold}{OT1}{cmss}{bx}{n}
  \ifAMStwofonts
    \ifCUPmtlplainloaded \else
      \DeclareSymbolFont{UPM}{U}{eur}{m}{n}
      \SetSymbolFont{UPM}{bold}{U}{eur}{b}{n}
      \DeclareSymbolFont{AMSa}{U}{msa}{m}{n}
      \DeclareMathSymbol{\upi}{0}{UPM}{"19}
      \DeclareMathSymbol{\umu}{0}{UPM}{"16}
      \DeclareMathSymbol{\upartial}{0}{UPM}{"40}
      \DeclareMathSymbol{\leqslant}{3}{AMSa}{"36}
      \DeclareMathSymbol{\geqslant}{3}{AMSa}{"3E}
    \fi
  \fi
\fi 

\ifCUPmtlplainloaded \else
  \ifAMStwofonts \else 
    \def\upi{\pi}
    \def\umu{\mu}
    \def\upartial{\partial}
  \fi
\fi

\title{The UV line spectrum of the soft X--ray 
transient XTE J1118+480: a CNO-processed core exposed}
\author[C.A. Haswell, R.I. Hynes, A.R. King, K. Schenker]
{C.A. Haswell$^1$, R.I. Hynes$^2$, A.R. King$^{3}$, K. Schenker$^{3}$\\
$^1$Department of Physics and Astronomy, Open University, Walton Hall,
Milton Keynes, MK7 6AA\\
$^2$Department of Physics and Astronomy, University of Southampton,
SO17~1BJ\\
$^3$Theoretical Astrophysics Group, University of Leicester,
Leicester, LE1~7RH\\
}

\date{Accepted
      Received }

\pagerange{\pageref{firstpage}--\pageref{lastpage}}
\pubyear{2001}

\begin{document}
\label{firstpage}

\maketitle
\begin{abstract}
We compare UV spectra of the recent soft X--ray transients \xs and
XTE\thinspace J1859+226.  The emission line strengths in \xs strongly
suggest that the accreting material has been CNO processed. We show
that this system must have come into contact with a secondary star of
about $1.5\msun$, and an orbital period $\sim 15$~hr, very close to
the bifurcation value at which nuclear and angular momentum loss
timescales are similar. Subsequent evolution to the current period
of 4.1~hr was driven by angular momentum loss. In passing through a
period of 7.75~hr the secondary star would have shown essentially
normal surface abundances. \xs could thus represent a slightly later
evolutionary stage of A0620--00.
We briefly discuss the broad Ly${\rm \alpha}$ absorption wings in \x .

\end{abstract}

\section{INTRODUCTION}
\label{sec:intro}
The recently-discovered soft X--ray transient (SXT) XTE~J1118+480 ( = KV UMa;
Remillard et al 2000) lies at high galactic latitude, close to the
Lockman Hole in the local ISM\@. The exceptionally low interstellar
absorption permits unprecedented wavelength coverage (Mauche et al
2000, Hynes et al 2000, McClintock et al 2001b, Chaty et al 2002). 

A low amplitude photometric modulation with period 4.1~hr was reported
by Cook et al (2000) and Uemura et al (2000).  Dubus et al (2000)
present spectroscopy showing an emission line radial velocity `S-wave' 
modulation at
a period close to this. 
This period was recently confirmed as orbital, and
a mass function of $f(M) \approx 6 \msun$ was 
measured 
 by McClintock et al
(2001a) and  Wagner et al (2001). 
Therefore the accretor clearly has to be
a black hole.

We present the UV line spectrum of \x, 
which we compare with that of another recent SXT, XTE\thinspace J1859+226.  
Full analyses of these HST spectra will be presented elsewhere (Hynes et al 2002a,b; Chaty et al 2002).
Here we focus on
the prior evolution of \x.

\section{OBSERVATIONS}

We obtained {\it HST}\/ spectra of XTE\,J1859+226 near the peak of
outburst on 1999 October 18 and 27 and November 6.  At each epoch
far-UV spectra were obtained using the low-resolution G140L grating
and wide (0.5\,arcsec) slit, yielding wavelength coverage
1150--1730\,\AA\ and resolution $\sim2.1$\,\AA\@.  We used the standard
pipeline data products except that the spectral extraction was done by
hand using the {\sc iraf} implementation of optimal extraction.  This
gave a significantly cleaner removal of geocoronal
emission lines (principally Ly $\alpha$ and O\,{\sc i} 1302\,\AA).
After extracting the one-dimensional spectra (with the pipeline
wavelength and flux calibrations), we resampled them onto a common
wavelength grid and took an exposure-time weighted average.

For XTE\,J1118+480, we obtained high-resolution (echelle) spectra
using the E140M grating and 0.2\,arcsec square aperture on 2000 April
8, 18 and 29, May 28, June 24, and July 8.  The pipeline extraction
was adequate, although a few high pixel values had to be replaced by
hand.  The extracted spectra were rebinned into 0.2\,\AA\ bins and
averaged with exposure-time weighting.
The process is clearly not perfect.  In particular, there are a number
of abrupt steps in the spectrum at longer wavelengths.  These
artefacts, at $\la 5$\,per cent level, arise from inconsistencies in
the relative flux calibration of adjacent echelle orders.

Figure 1 gives the UV spectra of \xs and XTE J1859+226. Both show
interstellar absorption features. In \xs these appear sharper because
the spectral resolution was higher.

\subsection{EMISSION LINES}
The difference in the equivalent widths of the emission lines in the
two systems is striking (Fig. 1, Table 1; Haswell et al 2000).  In XTE\thinspace
J1859+226 C\,{\textsc{iv}} is the strongest line, with N\,{\textsc{v}},
C\,{\textsc{iii}}, O\,{\textsc{v}}, Si\,{\textsc{iv}}, and
He\,{\textsc{ii}} also strongly present.  These emission line
strengths are roughly as expected for an X-ray bright LMXB (cf Sco
X-1, Kallman, Boroson, \& Vrtilek 1998).  Compared to cataclysmic
variables (CVs), the N\,{\textsc{v}}/C\,{\textsc{iv}} ratio is
relatively large, while Si\,{\textsc{iv}}/C\,{\textsc{iv}} is small, as
expected for a high ionisation parameter (Mauche, Lee \& Kallman
1997).  In contrast, for XTE\thinspace J1118+480, the carbon and
oxygen lines are undetectable, while the N\,{\textsc{v}} emission
appears enhanced.  A similar anomalous emission line spectrum is seen
in the magnetic CV AE Aqr, in which the N\,{\textsc{v}} is much
stronger than the C\,{\textsc{iv}} line (Jameson et al 1980, Eracleous
et al 1994).
\begin{figure*}
\begin{center}
\leavevmode
\psfig{angle=90,height=3.0in,width=7.25in,file=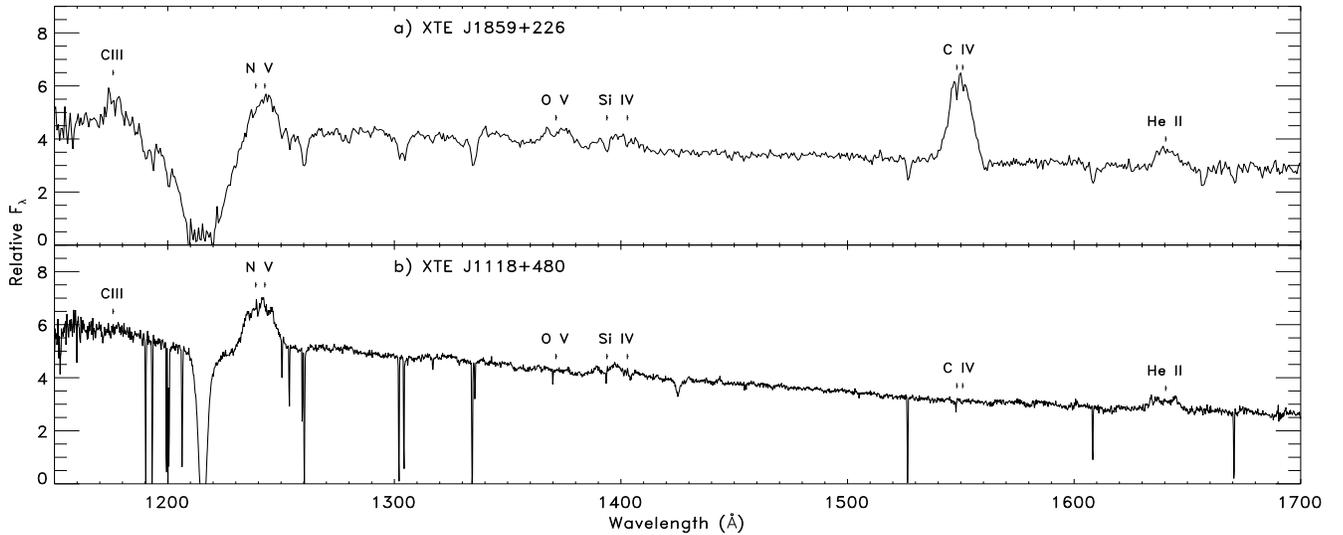}
\caption{HST/STIS UV emission line spectra. Upper panel: XTE J1859+480
shows normal line ratios, with C IV more prominent than N V. Lower
panel: XTE J1118+480: N V emission is prominent, but C IV and O V are
absent. 
}
\label{specfig2}
\end{center}
\end{figure*}

\begin{table*}
\begin{center}
{\leavevmode
\caption{Emission Line Characteristics}
\begin{tabular}{llrrr}
\hline

               &              & Line flux (erg\,s$^{-1}$\,cm$^{-2}$) & Equivalent width (\AA) & Gaussian FWHM (\AA) \\
\hline
XTE\,J1859+226 & N\,{\sc v}   & $(17.1 \pm 2.0)\times10^{-9}$                    & $ 8.0 \pm 1.4$         & 13                  \\   
               & O\,{\sc v}   & $( 4.5 \pm 0.2)\times10^{-9}$                    & $ 1.9 \pm 0.1$         & 15                  \\ 
               & Si\,{\sc iv} & $( 4.4 \pm 0.2)\times10^{-9}$                    & $ 2.0 \pm 0.1$         & 17                  \\
               & C\,{\sc iv}  & $(22.9 \pm 0.3)\times10^{-9}$                    & $11.8 \pm 0.2$         & 11                  \\
               & He\,{\sc ii} & $( 4.4 \pm 0.4)\times10^{-9}$                    & $ 2.4 \pm 0.3$         & 10                  \\
XTE\,J1118+480 & N\,{\sc v}   & $(24.0 \pm 0.5)\times10^{-13}$                   & $ 5.05\pm 0.14$        & 14                  \\   
               & N\,{\sc iv}   & $(\la 3 )\times10^{-14}$                   & $ \la 1.5     $        &                     \\   
               & O\,{\sc v}   & $(\la 2 )\times10^{-13}$                    & $ \la 0.6    $         &                   \\ 
               & Si\,{\sc iv} & $( 5.0 \pm 0.4)\times10^{-13}$                   & $ 1.29\pm 0.11$        & 18                  \\
               & C\,{\sc iv}  & $(\la 4)\times10^{-14}$                    & $\la 0.14 $         &                   \\
               & He\,{\sc ii} & $( 4.6 \pm 0.3)\times10^{-13}$                   & $ 1.74\pm 0.10$        & 15                  \\
\hline
\end{tabular}}

\end{center}
\end{table*}

The N\,{\textsc{iv}} 1718${\rm\AA}$ line would add a valuable additional
constraint on the interpretation of the line spectrum in terms of the
photo-ionisation conditions, so
we carefully assessed our \xs spectra 
to determine whether 
it
was present. Unfortunately our well-exposed E140M echelle
spectrum stops at 1710${\rm\AA}$ and our E230M echelle
spectrum suffers from shorter total
exposure and the low sensitivity of the NUV MAMA in this region.
1718${\rm\AA}$ falls right on the boundary between the two
STIS MAMA detectors, and consequently neither is optimised at
this wavelength.
We find no strong (${\rm \ga 3 \times 10^{-14}  erg~s^{-1}~cm^{-2}}$) line in the
spectrum, but this upper limit provides a weak constraint due to the
poor quality of our data at this wavelength.

\subsection{ABSORPTION LINES}
The Ly${\rm \alpha}$ absorption in \xs (Fig. 1) clearly has a sharp
core, which is probably interstellar, with very broad wings.  The
continuum slope is well constrained by the spectrum longwards of
$1280{\rm\AA}$. Extrapolating this continuum is suggestive of broad
Ly${\rm \alpha}$ extending from $\ga 1160{\rm\AA}$ to $\la
1280{\rm\AA}$.  
The combination of noise, N\,{\sc v} emission, and absorption features
hinder measurement of the Ly${\rm \alpha}$ wings, but they
definitely extend at least from $\sim 1180{\rm\AA}$ to $\sim 1230{\rm\AA}$.

The broad absorption may be damped Ly${\rm \alpha}$ from
the H\,{\textsc{i}} in our galaxy (Bowen, priv.\,comm.).
If we assume, instead, that it is due to absorbing gas executing
Keplerian motion around the black hole, and take the
largest width suggested by our data, the full width
implies velocities of $\sim 0.05~c$ and 
absorbing material at distances as close as $R_{\rm in} \sim 200
R_{\rm Sch}$.  Taking the securely determined line width
would imply absorption from Keplerian material at $R_{\rm in} \sim 500 R_{\rm Sch}$.

An independent line of argument based on the shape of the
spectral energy distribution, and dependent on the
value of the neutral hydrogen column density, $N_{\rm H}$,
has been used to  estimate the disc inner radius (Hynes et al 2000, 
McClintock et al 2001, Chaty et al 2002).
Hynes et al (2000) estimate $N_{\rm
H} = 0.75 \times 10^{20} {\rm cm}^{-2}$, a choice which
suggests the disc is terminated at $\ga
10^{3} R_{\rm Sch}$.  This is inconsistent with the
Keplerian interpretation of the broad Ly${\rm \alpha}$ absorption.
Alternatively, informed by the {\it Chandra}\/ data,
McClintock et al (2001) suggest that $N_{\rm H}$ could
be as high as $1.3 \times 10^{20} {\rm cm}^{-2}$, with 
$ R_{\rm in} \ga 65 R_{\rm Sch}$.
Chaty et al (2002) adopt $N_{\rm
H} = 1.0 \times 10^{20} {\rm cm}^{-2}$,
which leads to $ R_{\rm in} \sim 350 R_{\rm Sch}$.
The latter two values are
consistent with the Keplerian interpretation of the broad Ly${\rm \alpha}$ absorption.

The feature near ${\rm 1425\AA}$ does not
have an obvious identification and we believe it may be spurious, although
it does not correspond to any known instrumental artefacts (Sahu 2001
priv. comm.).  This issue is under investigation.

\section{IONISATION DIFFERENCES VERSUS ABUNDANCE ANOMALIES}

The C\,{\textsc{iv}} 1550${\rm \AA}$ and the N\,{\textsc{v}}
1240${\rm\AA}$ lines are both resonance lines of lithium-like ions,
and are produced under essentially the same physical conditions.
Kallman and McCray (1982, hereafter KM82) present theoretical models
of compact X-ray sources
predicting the ionisation structure expected in a wide variety of
astrophysical situations, including 
galactic X-ray binaries.
KM82 present 
8 distinct models; in each one the source
luminosity, $L$, spectral shape, and gas density, $n$, are fixed. For each model
the output includes the dependence on ionisation
parameter, 
\begin{equation}
\xi = {{L} \over {nR^2}},
\end{equation}
(where 
$R$ is the distance from the 
central source)
of
the abundances 
(as a fraction of the total elemental abundance)
of the ions of common elements. 

KM82 consider
(i) three species 
detected 
in both \xs and XTE\thinspace J1859+226: He\,{\textsc{ii}},
Si\,{\textsc{iv}}, and N\,{\textsc{v}};
and (ii) three species
detected only in the spectrum of
XTE\thinspace J1859+226:
C\,{\textsc{iv}}, C\,{\textsc{iii}}, and O\,{\textsc{v}}.
Strikingly
similar 
ionisation parameters were required for the two sets of lines
in all models. 
In almost all cases the presence of
N\,{\textsc{v}}  places the  tightest constraint on $\xi$, and 
almost invariably this is
encompassed within the range producing C\,{\textsc{iv}}.
In the minority of
cases where there exists an ionisation parameter which produces N\,{\textsc{v}}
and not C\,{\textsc{iv}}, this range also produces O\,{\textsc{v}}.
Consequently, KM82 give no
set of photoionisation parameters which would predict production of
N\,{\textsc{v}} in \xs
while suppressing the C\,{\textsc{iv}} and O\,{\textsc{v}}.

The X-ray spectrum of \xs is clearly different from that of XTE\thinspace
J1859+226: the former is an extended power-law
(Hynes et al 2000, McClintock et al 2001, Frontera et al 2001, 
Chaty et al 2002)
with a cut-off at $\sim 100$~keV
indicative of the low-hard state, 
and often attributed to an advection-dominated accretion flow (ADAF;
e.g. Esin et al 2001);
while
in the early stages of the outburst 
the latter was dominated by 
the soft thermal black-body disk spectrum
which turned over by $\sim 4$~keV
(Hynes et al 2002a),
typical of the high-soft state.
Ho et al (2000) note
that the characteristically harder ionising spectrum of an ADAF
lowers the effective ionisation parameter, and hence favours
the production of lower ionisation species.  Hence, if anything,
the differences in the X-ray spectra should cause 
C\,{\textsc{iv}} to be relatively prominent (compared with
N\,{\textsc{v}}) in \x , rather than
being suppressed as we observe.

Hamann and Ferland (1992) used the observed C\,{\textsc{iv}}
1550${\rm \AA}$ and N\,{\textsc{v}} 1240${\rm\AA}$ lines
to estimate the N/C abundance ratio in high redshift QSOs.  Their
photionisation calculations show that for a broad
range of ionisation parameters 
the N\,{\textsc{v}} 1240${\rm\AA}$ /
C\,{\textsc{iv}}
1550${\rm \AA}$ line ratio is lower at lower metalicities
even when the N/C abundance ratio is kept constant. This means
that if anything we should expect the N\,{\textsc{v}} 1240${\rm\AA}$
line to be suppressed relative to C\,{\textsc{iv}}
1550${\rm \AA}$ in \xs which is a halo object, and consequently
might be expected to have a lower metalicity than that of
XTE\thinspace J1859+226.

A definitive abundance analysis for \xs would require exact knowledge
of the geometry, densities, and ionising spectrum in the regions
emitting the UV lines.  While Hynes et al (2000, 2002b) gives
some information on the geometry, and Chaty et al. (2002) gives
good coverage of the X-ray spectrum, the EUV spectrum remains
open to significant uncertainty (compare the range of dereddened
EUV spectra in Chaty et al 2002, McClintock et al 2001, Esin et al
2001, Hynes et al 2000, Merloni et al 2000).  Furthermore the range
of densities
present in the UV line emitting gas is difficult to determine. 
Consequently a quantitative abundance determination
of the type carried out in stellar photospheres, where the physical 
conditions are well-known, is not possible.

Hence it is impossible to 
rule out definitively a contrived spectrum and gas distribution
which would
produce the set (i) lines while suppressing the
set (ii) lines. However as
\begin{itemize}
\item
{
set (i) and set (ii) ions require essentially identical
ranges of $\xi$
}
\item
{
the ionising spectrum  of \xs might be expected to favour 
C\,{\textsc{iv}} relative to
N\,{\textsc{v}}
}
\item
{
the lower metallicity expected for \xs might be expected to  boost
the C\,{\textsc{iv}}1550${\rm \AA}$ / N\,{\textsc{v}} 1240${\rm\AA}$
line ratio
}
\end{itemize}
by far the simplest explanation of our observed UV line spectra is
that
a substantial underabundance of carbon is present in the surface layers
of the mass donor star in \x .

In XTE\thinspace J1118+480 (as in AE Aqr), the underabundance of
carbon compared to nitrogen strongly suggests that the material in the
accretion flow is substantially CNO processed (Clayton 1983).
If the CNO bi--cycle achieves equilibrium, it converts most CNO nuclei
into $^{14}$N.  At the lower temperatures typical in $\sim 1.5 \,
\msun$ MS stars (which we shall suggest as the progenitor of the
companion in XTE\thinspace J1118+480) this conversion is much more at
the expense of $^{12}$C than $^{16}$O, and in fact $^{17}$O increases.
Hence the oxygen abundance is not expected to decrease by much in
XTE\thinspace J1118+480 
and this argument does not of itself explain the
observed weakness of O\,{\textsc{v}} 1371${\rm \AA}$. However
C\,{\textsc{iv}}1550${\rm \AA}$, N\,{\textsc{v}} 1240${\rm\AA}$ are
resonance lines, whose ratio must be regarded as a robust indicator of
conditions in the line emitting gas, whereas O\,{\textsc{v}} 1371${\rm
\AA}$ is a subordinate line, which may not be formed efficiently in a
low density photoionized gas. We thus regard the spectrum of
XTE\thinspace J1118+480 as indicating CNO processing.


\section{THE EVOLUTION OF XTE~J1118+480}

The spectrum shown in Figure 1 implies that the companion star in \xs
must be partially nuclear--evolved and have lost its outer layers,
exposing inner layers which have been mixed with CNO--processed
material from the central nuclear--burning region.
\begin{figure}
\begin{center}
\psfig{height=3.5in,width=3.5in,file=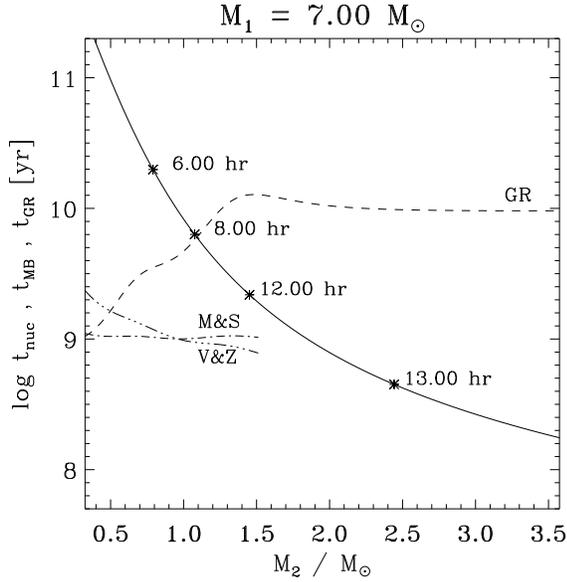}
\caption{Comparison of timescales for a $7 \, \msun$ black hole
primary. The full curve gives the nuclear timescale as a function of
the secondary mass. 
At various points the orbital period for a ZAMS star filling its
Roche lobe is indicated. 
The other 3 curves show the various relevant angular momentum loss
timescales $(-{\rm d} \ln J/{\rm d} t)^{-1}$: for gravitational
radiation (GR, dashed) and magnetic braking according to Mestel \&
Spruit (M\&S, dash-dotted) and Verbunt \& Zwaan (V\&Z,
dash-triple-dotted), all in the version of Kolb (1992)
Magnetic braking is assumed to be quenched in stars which have no
convective envelopes, i.e for $M_2 \protect\ga 1.5 \, \msun$.}
\label{taunuc}
\end{center}
\end{figure}

\begin{figure}
\begin{center}
\psfig{height=3.5in,width=3.5in,file=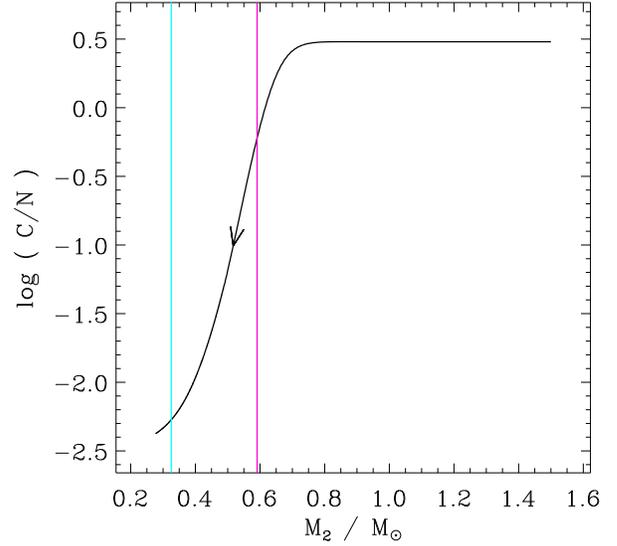}
\caption{Evolution of the surface abundance ratio C/N for a sequence
beginning mass transfer from a $1.5 \, \msun$ main sequence star on to
a $7 \, \msun$ black hole. At the turn--on period of 15~hr the core
hydrogen fraction had already been reduced to 28 \%. The current
period of \xs is reached at a mass of $0.33 \, \msun$, indicated by
the left vertical line, while the other near $0.6 \, \msun$ shows the
period of A0620-00. The transferred mass has been accreted by the
black hole which has grown beyond $8 \, \msun$.}
\label{surfratio}
\end{center}
\end{figure}

\begin{figure}
\begin{center}
\psfig{height=3.5in,width=3.5in,file=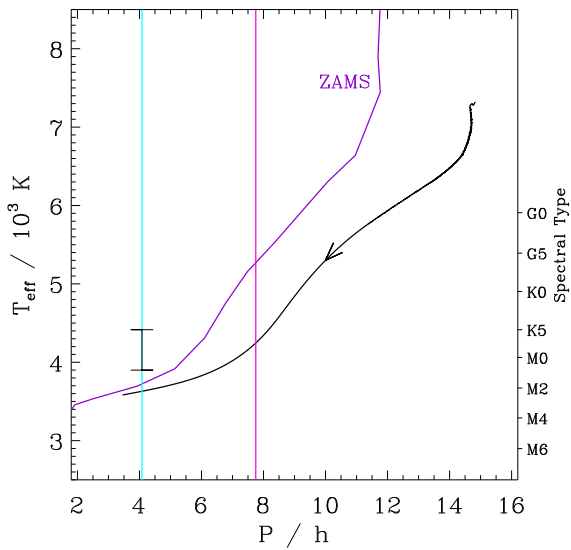}
\caption{Evolution of the effective temperature and spectral type with
orbital period for the sequence in Fig.~3 (lower curve), and the locus
of models with the donor on the ZAMS\@. 
The black hole mass is assumed to be $7 \, \msun$. Vertical lines indicate the
orbital period of A0620-00 and \xs respectively.
Error bars on the former mark the range of spectral type 
given by McClintock et
al (2000).} 
\label{xte_tep}
\end{center}
\end{figure}

Mass transfer must therefore have been initiated from a somewhat
evolved and sufficiently massive donor of $M_{2i}\ga 1.5\msun$, and
thus with an initial period $P_i \ga 12$~hr. The main difficulty in
understanding the current status of \xs is to explain its observed 
short orbital period; in general, significant nuclear evolution is associated
with a period {\it increase}, rather than the inferred decrease.

There are two ways in which this decrease could have occurred:

(i) the system came into contact with initial secondary/primary mass
ratio $q_i = M_{2i}/M_{1i} \ga 1$ and underwent a phase of
thermal--timescale mass transfer until $q \la 1$.  At the end of this
phase the period would have either decreased or not increased
greatly. Subsequent orbital angular momentum loss could then pull the
system in to the observed $P_{\rm orb} = 4.1$~hr. This case is similar
to the likely evolution of AE~Aqr (Schenker et al., 2002a).

(ii) the system came into contact with $q_i < 1$ already, and
subsequent evolution was driven by angular momentum loss towards
shorter periods.  This case resembles that proposed for A0620--00 by
de Kool et al (1983), with a severe additional constraint: the donor
must be sufficiently evolved to provide the observed surface abundance
ratios. This requires a near--equality of the nuclear and orbital
angular momentum loss timescales $t_{\rm nuc}, t_{\rm AML}$ (cf
Fig.~2). Put another way, $P_i$ must have been very close to the
`bifurcation period' defined by Pylyser \& Savonije (1988).

Case (i) above can be ruled out by the current system parameters. At
the end of thermal timescale mass transfer we must have $q \la 1$,
which would require a donor mass $M_{2} \sim 6\msun$ and an orbital
period $\sim 15$~hr. But for such a system $t_{\rm nuc} << t_{\rm AML}$
(cf Fig. 2), implying evolution to {\it longer}\/ periods, in stark
contrast to the current 4.1~hr. Put another way, there is an upper
limit on the binary mass $M = (1+q^{-1})M_2$ for case (i) evolution to give
nuclear--evolved donors at short orbital periods. The near equality
$t_{\rm nuc}\sim t_{\rm AML}$ can only realistically hold if magnetic
braking dominates angular momentum loss, so we require $M_2 \la
1.5\msun$ at the end of the thermal--timescale episode. But since $q
\sim 1$ there, we must have total binary mass $M \la 3\msun$ in case
(i) evolutions.

We are therefore left only with case (ii) above.  The secondary mass
when the system came into contact is now constrained to be close to
$1.5\msun$: CNO processing excludes significantly lower masses, while
the requirement for $t_{\rm AML} \approx t_{\rm MB} > t_{\rm nuc}$
excludes significantly higher masses.  Figs.~3, 4 show the evolution
of such a system: a $1.5 \, \msun$ MS star was allowed to evolve and
fill its Roche lobe in a 15~hr binary with a $7 \, \msun$ BH\@.  The
predicted mass transfer rate throughout the evolution, including its
current value $3 \times 10^{-10} \, \msun {\rm yr}^{-1}$, shows that
the system would indeed have appeared as a soft X--ray transient,
according to the irradiated--disc criterion for black--hole systems
given by King, Kolb \& Szuszkiewicz (1997).
It should be noted that the change in C/N is mostly due to a drastic
depletion of $^{12}$C, supplemented by a more modest increase in
$^{14}$N. The total O abundance at the surface has only decreased very
little; in fact the O/N ratio is reduced by only a factor of 3 to 5 at
most, predominantly due to the increase of N.
Interestingly, this model also passes through $P_{\rm orb} = 7.75 \,
{\rm hr}$ at around $M_2 \approx 0.6 \, \msun$, showing much weaker N
enhancements (i.e.\  almost normal abundances, cf Fig.~3).  Thus it
can also be considered a reasonable model for A0620--00. A subsequent
paper (Schenker et al., 2002b) will explore this and related
evolutions in detail. 

Finally we can compare further properties of the secondary in our
model to observations: Figure 4 shows the evolution of effective
temperature with orbital period, together with a simple mapping to
spectral types. 
The procedure is based on a conversion of observed colours to spectral
types (Beuermann et al., 1998) and a set of non-grey stellar
atmospheres (Hauschildt et al., 1999) providing the colours for each
set of stellar boundary conditions. For solar metallicity the
resulting SpT turns out to be only a function of effective temperature 
with a very weak dependence on surface gravity.
However this mapping should only be considered a first estimate,
as the evolutionary code in its current form still uses grey
atmospheres. Furthermore the whole conversion is based on an observed
set of unevolved stars and theoretical ZAMS models, i.e.\ no fully 
self-consistent application to a partially evolved MS star such as
shown in Fig.~4 is possible for the time being.
Allowing for the uncertainties described above, 
our model may be slightly too cool (M2 rather than K5-M1 as mentioned
in McClintock et al (2001), or K7-M0 by Wagner et al (2000).  
Similar evolutionary tracks of strongly evolved 
CVs are known to get hotter again (Baraffe \& Kolb, 2000) at short
periods, so a confirmed secondary spectral type would provide insight
as to the state of nuclear evolution in the donor star of \x. Figure 4
shows that in any case that the donor will currently
appear to be close to the ZAMS, and thus that the distance estimate
derived by McClintock et al (2000) on this basis is likely to be
quite good.

\section{Conclusions}
We have compared the UV spectra of \xs and XTE\thinspace J1859+226. The
former shows strong evidence of CNO processing, which tightly
constrains the evolution of the system. \xs must have first reached
contact with the donor star having a significantly nuclear--evolved
core, at a period where nuclear and angular momentum loss timescales
were comparable. This in turn constrains the end point of the
earlier common--envelope phase: 
immediately after the 
common--envelope phase the system had a wide enough separation
that significant nuclear evolution could occur before contact was
achieved. In a future paper we will investigate such evolutions
systematically.

\section{Acknowledgments}
CAH and RIH gratefully acknowledge the superb support of Tony Roman,
Kailash Sahu, and all involved in the implementation of our
time-critical {\it HST}\/ observations.  Support for proposals
GO-08245 and GO-08647 was provided by NASA through a grant from the
Space Telescope Science Institute, which is operated by the
Association of Universities for Research in Astronomy, Inc., under
NASA contract NAS5-26555. This work was supported by the Leverhulme
Trust F/00-180/A.  Theoretical astrophysics research at Leicester is
supported by a PPARC rolling grant. We thank Christian Knigge for
discussion about the nature of the ${\rm 1425\AA}$ feature, and an 
anonymous referee for useful comments.

\label{lastpage}

\end{document}